# STRATEGIC ALIGNMENT BETWEEN IT FLEXIBILITY AND DYNAMIC CAPABILITIES: AN EMPIRICAL INVESTIGATION

Van de Wetering, Rogier, Open University, the Netherlands, rogier.vandewetering@ou.nl

Mikalef, Patrick, Norwegian University of Science and Technology, Trondheim, Norway, patrick.mikalef@ntnu.no

Pateli, Adamantia, Ionian University, Corfu, Greece, pateli@ionio.gr

## Abstract

*Dynamic capabilities theory (DCT) emerged as a leading framework in the process of value creation for firms. Its core notion complements the premise of the resource-based view of the firm and is considered an important theoretical and management framework in modern information systems research. However, despite DCT's significant contributions, its strength and core focus are essentially in its use for historical firm performance explanation. Furthermore, valuable contributions have been made by several researchers in order to extend the DCT to fit the constantly changing IT environments and other imperative drivers for competitive performance. However, no DCT extension has been developed which allows firms to integrally assess their current state of maturity in order to derive imperative steps for further performance enhancements. In light of empirical advancement, this paper aims to develop a strategic alignment model for IT flexibility and dynamic capabilities and empirically validates proposed hypotheses using correlation and regression analyses on a large data sample of 322 international firms. We conjecture that the combined synergetic effect of the underlying dimensions of a firm's IT flexibility architecture and dynamic capabilities enables organizations to cope with changing environmental conditions and drive competitive firm performance. Findings of this study suggest that there is a significant positive relationship between firms' degree of strategic alignment—defined as the degree of balance between all dimensions—and competitive firm performance. Strategic alignment can, therefore, be seen as an important condition that significantly influences a firm's competitive advantage in constantly changing environments. The proposed framework helps firms assess and improve their maturity and alignment of IT flexibility and dynamic capabilities. We conclude with a discussion, suggestions for future research and managerial implications are also discussed.*

## 1 Introduction

Modern firms across a wide range of industries and sectors are constantly seeking competitive potential in order to improve customer efficiency, firm effectiveness, and transform the organization toward a sustainable and evolutionary fit business model driven by organization-wide innovations (Dao, Langella, & Carbo, 2011; Hanelt, Busse, & Kolbe, 2016; Seidel, Recker, & Vom Brocke, 2013). The adoption, effective use, and alignment of information systems and information technology (IS/IT) are critical in this respect (Hanelt et al., 2016; Malhotra, Melville, & Watson, 2013; Melville, Kraemer, & Gurbaxani, 2004; Wade & Hulland, 2004). However, various scholars have argued that firms need to deal with various complexities arising from aligning business operations and IS/IT domains (Sabherwal, Hirschheim, & Goles, 2001; Wegmann, 2002), while also taking into account the dynamics of the environment (Eisenhardt & Martin, 2000; Teece, Pisano, & Shuen, 1997) and continuous organizational change (Brown & Eisenhardt, 1997). It has been well documented in the literature that this particular process complements the leveraging of idiosyncratic and intangible re-sources to build competences (Wernerfelt, 1984), i.e., the internal oriented resource-based view (RBV) of the firm. The dynamic capabilities theory (DCT) extended this rather static RBV (Teece et al., 1997) and embraced environmental influences and market dynamism (Wang & Ahmed, 2007). Eisenhardt and Martin (2000) observed that dynamic capabilities can only be effective if they match the rate of environmental changes and not all enterprise-level responses to external stimuli are manifestations of dynamic capabilities (Winter, 2003). Moreover, Henderson and Venkatraman stressed that alignment, as a dynamic capability, is not an ad-hoc event, but rather a process of continuous adaption and change. As such, they argued that 'no single IT application—however sophisticated and state of the art it may be—could deliver a sustained competitive advantage' (Henderson & Venkatraman, 1993). It is also within this particular context that IT capabilities and scalable enterprise IS/IT infrastructures have been proposed as a means to achieve a competitive edge (Duncan, 1995; Kim, Shin, Kim, & Lee, 2011; Tiwana, Konsynski, & Bush, 2010).

IT capabilities have been viewed by past literature as complex, multidimensional constructs (Pavlou & El Sawy, 2006). As such, scholars and practitioners have used inconsistent conceptualizations of this term, while complementary perspectives that investigate the dynamics among different types of IT capabilities (e.g., IT flexibility, human and management capabilities), have been largely overlooked (Fink, 2011; Kim et al., 2011). Past literature, suggested that the unique characteristics of an IT infrastructure determine the value of that infrastructure to organizations (Byrd & Turner, 2000). Moreover, a firm's IT flexibility is regarded as a critical aspect of an IT infrastructure that can potentially influence a firm's ability to use and reconfigure IT strategically (Bharadwaj, 2000; Bhatt & Grover, 2005; Ray, Muhanna, & Barney, 2005). Hence, we focus on the IT flexibility dimension of the IT capability, which is likely to help firms differentiate themselves from competition and drive competitive firm performance (Mikalef, Pateli, & Van de Wetering, 2016). In doing so, we build upon Simon's theory of near decomposability, i.e., his design principles for modular systems and loose coupling (Simon, 1965; Weick, 1976), which have also been linked to increased levels of strategic alignment under volatile circumstances that require agile and swift responses by the firm (Tallon & Pinsonneault, 2011). This demonstrates that a flexible IT infrastructure can facilitate a timely response in terms of IT-based competitive actions, geared towards sustained competitive advantage (Overby, Bharadwaj, & Sambamurthy, 2006). In this respect, the IT infrastructure is not only used to support current operations but is developed on the basis of constant adaptations, or as referred to, a platform for digital options (Overby et al., 2006; Sambamurthy, Bharadwaj, & Grover, 2003).

Although IT flexibility may to some extent strengthen a firm's armory of digital options, it is conceivable that this imperative dimension in isolation may not be sufficient to drive firm performance. This relates well to the principles of efficacious IS/IT adaptation and coevolution: "*having none of them is a disaster; having all particularly feeds adaptive and synergetic success*" (Benbya & McKelvey, 2006). This synergetic success especially fits the core concept of strategic alignment, i.e., equilibrium of different organizational dimensions, and external fit as strategy development that is based on environmental trends and changes (Chan & Reich, 2008; Henderson & Venkatraman, 1993; Van de Wetering & Batenburg, 2014) and also more recent work on facilitation of dynamic capabilities (Sher & Lee, 2004), co-evolutionary relationship between IT investments, capabilities and their ability to launch competitive actions (Patrick Mikalef et al., 2016; Sambamurthy et al., 2003). Hence, we foresee synergies arising from IT flexibility and complementary organizational capabilities in strengthening a firm's armory to improve competitive firm performance (Bhatt & Grover, 2005; Lane, Salk, & Lyles, 2001; Roberts, Galluch, Dinger, & Grover, 2012; Wade & Hulland, 2004). We, therefore, posit that the combined synergetic effect of the underlying dimensions of IT flexibility architecture and dynamic capabilities enables organizations to cope with changing environmental conditions and drive competitive firm performance. Furthermore, we employ the basic thought that in order to truly understand the nature of dealing with changing environmental conditions and achieve competitive firm performance, a coherent framework is required that fits the diversity, interdependencies, and alignment of all involved organizational dimensions (Van de Wetering, 2016). Hence, we build upon shortcoming of extant literature and on the theoretical developments of the DCT, modular systems theory, and general IS/IT-alignment literature.

As such, the main goal of this paper is to examine if the strategic alignment of both IT architecture flexibility and dynamic capabilities contributes to higher levels of competitive performance. Doing so, we develop a strategic alignment model for IT flexibility and dynamic capabilities and empirically validate this model at international firms. Hence, in this article we address the following research question:

> *"To what extent does alignment of key dimensions of both IT flexibility and dynamic capabilities enhance competitive firm performance?"*

This research is valuable since outcomes will support firms to utilize their current and future IT infrastructure capabilities to support evolutionary fitness with the external environment (Helfat & Peteraf, 2009). The remainder of the article is structured as follows. In section 2, we begin with a brief review of the DCT, IT flexibility, and IS/IT alignment literature, in order to describe the context of this study. Section 3 introduces the conceptual model and presents hypotheses. In the last sections of the paper, results are presented and discussed, inherent limitations are identified and future research opportunities are addressed.

## 2 Theoretical development and conceptual framework

### 2.1 Dynamic capabilities

The RBV of the firm (Barney, 1991; Wernerfelt, 1984) has been one of the most influential theoretical frameworks for understanding how firms attain competitive performance gains as a result of their resources and capabilities. Nevertheless, the theory does not place extensive emphasis on the locus of long-term competitive advantage in dynamic markets and changing business conditions (Teece et al., 1997). This theoretical perspective has been subsequently extended to dynamic markets (Teece et al., 1997; Wilden & Gudergan, 2015) in an attempt to explain how and why certain firms have a competitive advantage in situations of rapid and unpredictable change (Eisenhardt & Martin, 2000). The DCT has been developed by identifying the main routines that allow a firm to change and reconfigure when the opportunity or need arises (Eisenhardt & Martin, 2000; Helfat & Peteraf, 2009; Wang & Ahmed, 2007). Past empirical studies have relied on the definitions of Teece et al. (Teece et al., 1997) (reconfiguring, learning, integrating, and coordinating), and Teece (Teece, 2007) (sensing the environment to seize opportunities and reconfigure assets) in order to isolate these routines. Following the approach described above, existing literature suggests that dynamic capabilities comprise of the following routines: (i) sensing, (ii) coordinating, (iii) learning, (iv) integrating, and (v) reconfiguring (Mikalef et al., 2016; Pavlou & El Sawy, 2011; Protogerou, Caloghirou, & Lioukas, 2012). IS and management scholars have coined the term IT-(enabled) capabilities in an attempt to measure a firm's proficiency in exploiting it's IS/IT assets, competences and capabilities (El Sawy & Pavlou, 2008). Within mainstream management and IS research, an IT capability is not so much a specific set of technological functionalities as it is an enterprise-wide capability to leverage technology to differentiate from competition and foster agility (Lu & Ramamurthy, 2011). However, IT capabilities are merely conceptualized as an aggregation of IT resources and IT competencies in the vast majority of empirical studies (Wade & Hulland, 2004). Synthesizing from the above, the DCT is therefore considered an appropriate framework to explain how firms can differentiate and compete in a turbulent environment, taking into account that they must evolve and co-evolutionary reconfigure their (IS/IT) operations in order to remain competitive.

### 2.2 IT flexibility and modular systems theory

Modular systems design dates back to Simon's theory of near decomposability, i.e., his design principles for modular systems and 'loose coupling' (Simon, 1965; Weick, 1976). This theory argues that complex systems consisting of modular, or else nearly decomposable subunits, tend to evolve faster, increase the rate of adaptive response and tune towards stable, self-generating configurations (Simon, 1965). In the same vein, Schilling (2000) proposed a general modular systems theory (GMST). GMST states that many systems opt towards modular forms in order to enable greater agility in end configurations. The general assumption is that many complex systems adapt or evolve in response to changes in their context, thus, increasing independency between sub-systems will lower the need for coordinated changes in others. A system may adapt purposefully, as when organizations alter themselves to better seek value (Kim & Pae, 2007). Modularity, as such, is a characteristic which largely determines the effectiveness in implementing continuous change, and is suggested to be an antecedent of dynamic capabilities (Pil & Cohen, 2006). Research on modularity in IS research has e.g., been examined as the flexibility of the IT architecture and the decentralization of the IT governance structure (Byrd & Turner, 2000), and has emerged as a *de facto* standard (Wilkinson, 2006) and key competitive priority in many organizational activities (Ray et al., 2005). IT flexibility can be regarded as a critical component to reconfigure IT strategically (Bharadwaj, 2000; Bhatt & Grover, 2005; Ray et al., 2005), an enabler of strategic alignment under circumstances that require agile and swift responses by the firm (Tallon & Pinsonneault, 2011), and a facilitator of IT-based competitive actions and a platform for digital options (Overby et al., 2006).

## 2.3 The concept of IS/IT-alignment

Investments in IS/IT, along with structured adoption and use, have been suggested to lead to multifactorial advantages and competitive gains for organizations in various industries (Tallon & Pinsonneault, 2011). These financial and non-financial gains include more efficient processes, reduction of costs, better deals with business partners, and less human errors amongst others (Devaraj & Kohli, 2003). Yet, despite heavy investments in IT, organizations quite often fail to achieve improvements in their organizational performance due to their inability to align IT with organizational needs. In general, this so-called 'productivity paradox' (Strassmann, 1990) has been greatly attributed to the lack of fit, or else alignment, between business strategy and internal resources including IT. Both in scientific literature and in practice, it is a well-known fact that achieving a state of IS/IT-alignment is a crucial step in order to leverage the maximum potential benefits (Brynjolfsson & Hitt, 2000; Gerow, Grover, Thatcher, & Roth, 2014; Henderson & Venkatraman, 1993). IS/IT-alignment has been a major concern for executives and IT practitioners for decades and refers to applying IS/IT in an appropriate and timely way, in harmony with business strategies, goals, and needs (Luftman & Kempaiah, 2007). Achieving IS/IT-alignment comes with various performance gains, including market growth, cost control, financial performance, increasing rates of innovation, and augmented reputation (Kearns & Lederer, 2003). The current literature points out that IS/IT-alignment remains a top priority for business and IT executives (Gerow et al., 2014). Following both recognized work and more recent studies (Avison, Jones, Powell, & Wilson, 2004; Gerow et al., 2014; Van de Wetering & Batenburg, 2014) we argue that little scientific knowledge is available about the underlying theoretical mechanisms that govern competitive firm performance and how BITA contributes to this as an antecedent (Van de Wetering, 2016).

# 3 Conceptual model and hypotheses

## 3.1 Underpinnings of the conceptual model

### 3.1.1 Dynamic capabilities

For the purpose of this study, we adopted five elementary dimensions of dynamic capabilities, i.e.: (1) sensing, (2) coordinating, (3) learning, (4) integrating, and (5) reconfiguring routines (Mikalef et al., 2016; Pavlou & El Sawy, 2011; Protogerou et al., 2012). Since the construct of dynamic capabilities is a novel one, past empirical literature was referenced to incorporate specific measures (Mikalef &

Pateli, 2016). Literature from the areas of strategic management, information systems, and organizational science literature was used to formulate items as presented in Appendix A.

### 3.1.2 IT flexibility

We define IT flexibility as the degree of decomposition of an organization's IT portfolio into loosely coupled subsystems that communicate through standardized interfaces. According to the definition of (Byrd & Turner, 2000), the degree of shareability and reusability of an IT architecture define what is known as IT flexibility. Each of the dimensions that comprise IT flexibility is measured based on past empirical work (Byrd & Turner, 2000; Tafti, Mithas, & Krishnan, 2013; Tiwana et al., 2010). Hence, we identify the following elementary dimensions for IT flexibility, i.e.: (1) loose coupling, (2) standardization, (3) transparency and (4) scalability.

### 3.1.3 Competitive performance

Competitive performance refers to the degree to which a firm performs better than its key competitors (Rai & Tang, 2010). This definition is in accordance with studies evaluating IS/IT performance from a rich and diverse understanding of outcomes, i.e., a multi-factorial perspective (Hitt & Brynjolfsson, 1996; Kohli & Devaraj, 2003; Van de Wetering & Batenburg, 2014). In the context of alignment, competitive advantage in comparison with competitors should not solely be translated into hard to quantify economic values (Patrick Mikalef, Pateli, Batenburg, & Van de Wetering, 2013). As such, the performance construct we employ in this study enables a more diverse understanding of outcomes from various perspectives. Following validated performance work (Li & Zhou, 2010; Liu, Ke, Wei, & Hua, 2013; Rai & Tang, 2010) we identify the following elementary items, i.e.: (1) Return on investment (ROI), (2) Profits as percentage of sales, (3) Decreasing product or service delivery cycle time, (4) Rapid response to market demand, (4) rapid confirmation of customer orders, (5) Increasing customer satisfaction, (6) In profit growth rates, (7) In reducing operating costs, (8) Providing better product and service quality, and (9) Increasing our market share. Respondents had to score all competitive performance items using a Likert scale from 1 – much weaker than the competition to 7 – much stronger than the competition.

## 3.2 Strategic alignment model for IT flexibility and dynamic capabilities

Strategic alignment or fit has been conceptualized in different ways by many scholars and has received substantial attention in the literature (Bergeron, Raymond, & Rivard, 2004; Chan & Reich, 2007). This study employs the 'profile deviation' perspective, subsequently defines a specified (alignment) profile and proposes that this multidimensional profile will be positively related to competitive firm performance (Chan & Reich, 2008; Venkatraman, 1989). Under this profile deviation conceptualization, we deduce from extant literature an ideal alignment profile of dimensions that facilitate alignment and hence contribute to competitive firm performance (Batenburg, Helms, & Versendaal, 2006; Batenburg & Versendaal, 2004; Van de Wetering et al., 2011). Applied to the current context, individual elements of the core dimension should thus be aligned, in order to achieve optimal competitive performance. Our conceptual model is illustrated in Figure 1. Depicted are (1) dimension of IT flexibility and dynamic capabilities (to be aligned) on the horizontal axis and (2) the dynamics of alignment, i.e., represented by a balance between the connected dots. The conceptual model depicts an illustrative outcome of such a firm assessment. The average scores on a single dimension, i.e., maturity score on each dimension, are represented by a single dot. The concept of IS/IT maturity is an essential element of our strategic alignment model, and in essence provides insight into the structure of elements that represent process effectiveness of IS/IT in organizations and give guidance through this evolutionary process by incorporating formality into the improvement activities (Jiang, Klein, & Shepherd, 2001; Mettler, Rohner, & Winter, 2010; Sledgianowski, Luftman, & Reilly, 2006). Hence, we define 'optimal' alignment, for both the IT flexibility and dynamic capabilities part of the model, if the connected dots form a vertical line in the different dimensions (Van de Wetering et al., 2011). Drawing on our theoretical perspective, the conceptual model posits that alignment of all dimensions is positively related to competitive firm performance. To be more specific, our model is built on two foundations, i.e., I. the alignment of dynamic capabilities and II. additional contributions to competitive performance through the 'simultaneous' alignment of all IT flexibility dimensions. Therefore, we embrace the concept of decomposition into the core conceptual components (Overby et al., 2006).

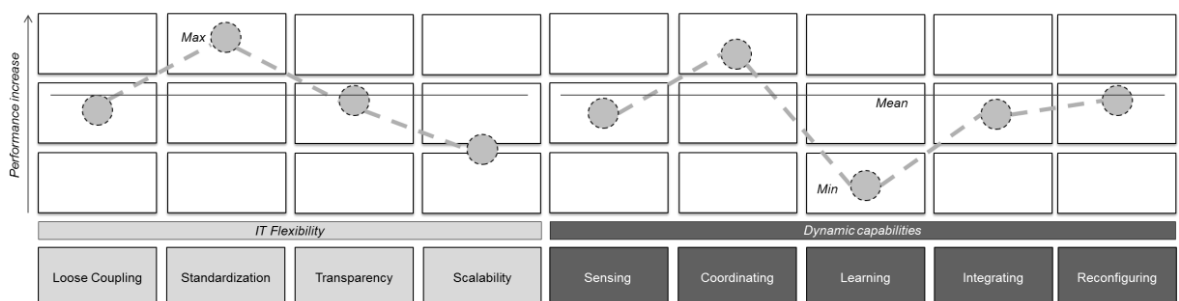

*Figure 1.*        *Strategic alignment model for IT flexibility and dynamic capabilities (with illustrative 'simultaneous' outcomes)*

Based on the research question and the proposed alignment model, we define the following main proposition:

> **Main proposition.** *Alignment of IT flexibility and dynamic capability dimensions has a positive impact on competitive firm performance.*

Synthesizing from general IS/IT alignment and maturity literature, performance is improved and productivity gains are achieved with the progression towards higher levels of alignment (cf. Galin & Avrahami, 2005). It is even argued that synergies arise from IT capabilities and complementary dynamic capabilities (Roberts et al., 2012). Thus, we want to understand how synergetic alignment effects of the underlying dimensions of IT flexibility architecture, dynamic capabilities and their combined (simultaneous) effect enables competitive firm performance. Based on the above and underpinned by profile deviation perspective, we define the following hypotheses:

> **Hypothesis 1.** *Alignment of IT flexibility dimensions is positively associated with competitive firm performance.*

> **Hypothesis 2**. *Alignment of dynamic capability dimensions is positively associated with competitive firm performance.*

Within this alignment conceptualization, we differentiate between 'simultaneous' and 'strategic' alignment profiles. Simultaneous alignment uses two separate predictor variables, i.e., IT flexibility and dynamic capabilities whereas integral alignment is operationalized using a single predictor variable covering all underlying IT flexibility and dynamic capability dimensions. With the strategic perspective, we foresee and hypothesize that combinative alignment of the all underlying IT flexibility and organizational dynamic capabilities dimension together will even more strongly drive a firm's competitive performance gains (Melville et al., 2004; Roberts et al., 2012; Wade & Hulland, 2004). Hence, we define:

> **Hypothesis 3**. *'Simultaneous' alignment of IT flexibility and dynamic capability dimensions is 'more strongly' associated with competitive firm performance than alignment of IT flexibility or dynamic capabilities in isolation.*

> **Hypothesis 4a**. *'Strategic' alignment of IT flexibility and dynamic capability dimensions is positively associated with competitive firm performance.*

> **Hypothesis 4b**. *'Strategic' alignment of IT flexibility and dynamic capability dimensions is more strongly associated with competitive firm performance than other alignment models.*

# 4   Methods

## 4.1   Data collection

To measure the previously mentioned concepts, a questionnaire was developed, that included 50 questions covering all relevant dimensions (Appendix A). All items used a Likert scale from 1 – strongly disagree to 7 – strongly agree. The applied survey has been pretested (Mikalef et al., 2016) and non-response bias actions were taken into account. The final survey was sent to key informants within firms, including Chief Information Officers (CIO), IT managers, Chief Technology Officers (CTO), enterprise architects, and Chief Executive Officers (CEO). In total 1500 firms were randomly selected from the ICAP business directory, comprising of firms from almost all industries and sectors. To assure a collective response, the instructions asked executives to consult other members of their firm for information they were not highly knowledgeable about. The duration of the data gathering process was approximately nine months (January 2015 – September 2015). In total, we incorporated 322 usable questionnaires yielding a valid response rate of 21.4%, which is consistent with comparable studies using key informant methodology (Capron & Mitchell, 2009). In order to control *ex-ante* for common method bias, respondents were assured that data collected would remain anonymous, and would be used solely for research purposes at an aggregate level. In addition, in order to control for common method variance *ex-post*, Harman's single factor test was performed, in which it was found that the majority of variance could not be attributed to one factor. The majority of responses were from consulting services (24%), high-tech (24%), financials (14%), consumer goods (10%), telecommunications (6%), industrials (6%), and consumer services (5%) industries. Less than 5% were obtained from the basic materials, healthcare, utilities, and oil & gas industries. The survey was in most cases completed by chief information officers (CIOs), chief executive officers (CEOs) and IT managers. In accordance with the EU Commission size-class recommendation (2003/361/EC), firms were grouped into large (38%), medium (20%), small (26%), and micro (16%).

## 4.2   Analysis and alignment calculations

### 4.2.1   Preliminary reliability and validity tests

Indicator and construct reliability and validity assessments were done through the use of SPSS v22. Reliability was gauged at the construct and item level. Construct reliability was established by examining that all Cronbach's Alpha values for both independent and dependent constructs were above the threshold of 0.70 (Nunally & Bernstein, 1978). Item reliability was assessed by examining if construct-to-item loadings were above the threshold of 0.70. Items with lower loadings were omitted from the measurement model. Convergent validity was also assessed (Fornell & Larcker, 1981); all items greatly exceeded required threshold values.

### 4.2.2   Operationalizing and formalizing alignment

For this paper, we define alignment as the degree of balance between all defined dimensions (Van de Wetering et al., 2011). This operationalization follows the core idea of Scheper's extension (Scheper, 2002) of the strategic alignment model of Turban (Turban, McLean, & Wetherbe, 1999), i.e., synchronizing or balance of all dimensions (as an ideal profile) will significantly contribute to the performance of an organization. Hence, alignment can be expressed within Figure 1 as a line connecting all (vertical) success dimension. There are different methods possible to calculate the differences between the maturity scores of the various dimension of IT Flexibility and dynamic capabilities. Here we follow specific procedures of Batenburg et al. (2004) by incorporating both mean scores ($\mu$) as well as the difference between the maximum and minimum maturity scores of the dimensions (for both IT flexibility and dynamic capabilities) as a measure for alignment. Hence, we simultaneously and integrally include both deviations as well as the means of the score array, or maturity. This mean score is then multiplied by the ratio of the minimum (MIN[$IT1..IT4$] and [$DC1..DC5$]) and the maximum score (MAX [$IT1..IT4$] and [$DC1..DC5$]) within that same array. Thus, alignment of both the IT flexibility and dynamic capabilities part of the model can be formalized as:

$$Alignment = \mu \text{ x } (Min/Max)^1$$

Following this logic, the 'smaller' the difference all dimensions, the 'better' the alignment between the dimensions. There are many alternatives to measure alignment, e.g., calculating the standard deviation, or selecting the minimum score (as the 'weakest link'). In practice, these alternative measurements for alignment strongly correlate with our alignment measurement. As for the competitive performance, we argue that they inter-correlate positively. We tested this using reliability analysis (Cronbach's alpha of 0.93). This indicates that this is a highly reliable scale. We, therefore, incorporate the mean score of the variables as a dependent variable in the model and hence the statistical analysis.

## 5    Results and hypotheses testing

### 5.1    Preliminary tests and descriptive statistics

We first describe the basic characteristics of our dependent and independent variables as explained in the previous sections. As can be seen in Table 1, respondents score on average (perceived performance) 4.77 on a scale of 1-7 with a standard deviation of 1.22.

|  | N | Mean | Std. Dev. | Zero-order correlation | Partial correlations |
|---|---|---|---|---|---|
| *Independent variables* |  |  |  |  |  |
| Alignment of IT flexibility | 322 | 3.86 | 1.41 | .478*** | .455*** |
| Alignment of DCs | 322 | 3.87 | 1.31 | .495*** | .449*** |
| Strategic alignment | 322 | 3.38 | 1.32 | .512*** | .471*** |
| *Dependent variable* |  |  |  |  |  |
| Competitive performance | 322 | 4.77 | 1.22 |  |  |

*Table 1.        Descriptive Statistics *** Correlation is significant at the 0.0001 level (2-tailed)*

Table 1 also outlines the preliminary test of our hypotheses. As such, we initially performed a standard Pearson correlation analysis in SPSS v22 and tested on two-tailed significance. A Pearson correlation analysis is, in essence, an adequate assessment method to explore our main proposition, i.e., alignment of IT flexibility and dynamic capability dimensions positively impacts competitive firm performance. In addition, we performed a partial correlation whilst controlling for 'size' of the organization and various environmental factors, i.e., 'dynamism', 'complexity' and 'hostility', which will be elaborated upon in section 5.2. Table 1 shows similar significant results of the two-tailed zero-order and partial correlation analyses for alignment of dynamic capabilities, alignment of IT flexibility dimensions as well as their integral (strategic alignment) combination. Therefore, hypotheses 1, 2 and 4a initially seem to be supported. We do see that correlation values decrease marginally if we control for size and environmental factors. Alignment of dynamic capabilities seems to be impacted most. In general, the proposition assumed that there would be a positive association between the independent variables and the dependent variable; empirical evidence supports this assumption. Results also show that the strategic alignment perspective, i.e., 'integral' alignment of both (1) dynamic capabilities and (II) IT flexibility dimensions, thus operationalized using a single predictor variable covering all underlying IT flexibility and dynamic capability dimensions, has a statistically stronger association with competitive firm performance than the other independent variables. This will be further investigated in the next section.

---

[1] Example calculation: Mean scores ($\mu$) are calculated for all the IT flexibility dimensions, i.e., Modularity, Transparency, Standardization and Scalability with respective scores: 7, 4, 5, 6 ($\mu$ = 5.5). Next, the ratio of the minimum MIN [4] and the maximum score MAX [7] is calculated, i.e., 4/7 = 0,57. Finally, we include both $\mu$ and the calculated ratio. Thus, alignment = $\mu$ x (Min/Max) = 5,5 x 0,57 = 3,14

## 5.2 Regression analyses

The normal P-P and scatter plots (figure 2) showed that our data is normally distributed (i.e., all residuals cluster around the 'line'), complies with the assumptions of homogeneity of variance (i.e., homoscedasticity) and linearity. Residual errors are in fact evenly distributed and not related to the value of the predicted value, suggesting that the relationship is, in fact, linear and the variance of *y* for each value of *x* is the same, this confirming the homoscedasticity assumption (Kachigan, 1991). We checked for univariate outliers using z-scores and all values were within acceptable range. In succession, we checked for multivariate outliers using Mahalanobis and Cook's distances. No influential outliers were detected. We also checked for multicollinearity using variance inflation factors (VIF). Results of the regression analysis are summarized in Table 2.

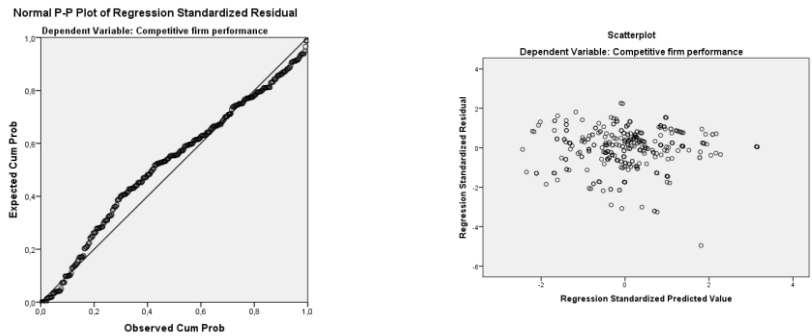

*Figure 2.    Normal P-P plot of regression standardized residual and residual scatter plot*

Like in the correlation analyses, we controlled for potential contextual influences, namely 'size' (based on the survey items: 'Indication of the size-class of your company'), and three potentially moderating variables concerning environmental uncertainty, i.e., 'dynamism', 'complexity' and 'hostility'[2] (Chen et al., 2014). These moderating effects have previously been proposed and empirically validated (Pavlou & El Sawy, 2006). Interaction terms were obtained by multiplying the standardized predictor variables with the standardized moderator variable. Following Baron and Kenny (1986) and Tallon and Pinsonneault (2011), we represented the moderating effect as an interaction between a focal predictor (alignment) and a factor that specifies the appropriate conditions for its operation (environmental factors). Therefore, we used the main and the interaction effect in the regressions, although the main (in this case conditional) effect of the included moderating variables need not be interpreted directly (Hayes, 2013). Moderation can be tested using regression analyses (Fairchild & MacKinnon, 2009). Hence, we used a hierarchical regression approach whereby we first added the control variable (size) and the moderating variables, then the main predictors and finally the specified interaction terms. Table 2 shows the final regression models.

Strategic alignment (model D) of both (1) dynamic capabilities and (II) IT flexibility dimensions (hypothesis 4a) significantly predicts competitive firm performance, $\beta = .495$, p< .0001. In addition, alignment also explained a significant proportion of variance in competitive performance, $R^2 = .300$.

---

[2] Hostility is represented by I) Scarce supply of labor, II) Scare supply of materials, III) Tough price competition, IV) Tough competition in product/service quality and V) tough competition in product/service differentiation (Cronbach's alpha = .67).

Hereby hypothesis *4a* is confirmed. We also run a regression analysis using both I) alignment of IT flexibility dimensions (model A) and II) alignment of dynamic capability dimensions (model B) as single predictors variables within the regression model in order to test hypotheses *1* and *2*. Also, we checked for their combined simultaneous effect (model C), hypothesis *3*. Model A significantly predicts competitive form performance, $\beta = .467$, p< .0001.

| Alignment models | N | $R^2$ | $R^2$ (Adj.) | $\beta$ | *t*-Value | VIF | *F*-scores |
|---|---|---|---|---|---|---|---|
| **Model A: IT flexibility alignment** | 322 | .286 | .267 | | | | 15.649 |
| *Size* | | | | -.014 | -.287 | 1.081 | |
| *Dynamism* | | | | -.091 | -1.753 | 1.193 | |
| *Complexity* | | | | -.025 | -.408 | 1.594 | |
| *Hostility* | | | | .248 | 4.304*** | 1.453 | |
| *Alignment of IT flexibility* | | | | .467 | 9.180*** | 1.132 | |
| *Dynamism x Alignment* | | | | -.085 | -1.536 | 1.337 | |
| *Complexity x Alignment* | | | | -.036 | -.611 | 1.486 | |
| *Hostility x Alignment* | | | | -.036 | -.611 | 1.486 | |
| **Model B: DCs alignment** | 322 | .278 | .260 | | | | 15.069 |
| *Size* | | | | .015 | .293 | 1.109 | |
| *Dynamism* | | | | -.089 | -1.686 | 1.218 | |
| *Complexity* | | | | .015 | .246 | 1.522 | |
| *Hostility* | | | | .159 | 2.637* | 1.585 | |
| *Alignment of DCs* | | | | .463 | 8.889*** | 1.175 | |
| *Dynamism x Alignment* | | | | -.072 | -1.230 | 1.472 | |
| *Complexity x Alignment* | | | | .069 | .996 | 2.066 | |
| *Hostility x Alignment* | | | | -.024 | -.360 | 1.976 | |
| **Model C: Simultaneous alignment** | 322 | .327 | .301 | | | | 12.518 |
| *Size* | | | | .016 | .320 | 1.132 | |
| *Dynamism* | | | | -.082 | -1.585 | 1.225 | |
| *Complexity* | | | | -.029 | -.480 | 1.629 | |
| *Hostility* | | | | .207 | 3.428** | 1.682 | |
| *Alignment of DCs* | | | | .272 | 4.017*** | 2.107 | |
| *Alignment of IT flexibility* | | | | .305 | 4.661*** | 1.968 | |
| *Dynamism x Alignment IT flex.* | | | | -.063 | -.901 | 2.270 | |
| *Complexity x Alignment IT flex.* | | | | -.003 | -.038 | 3.649 | |
| *Hostility x Alignment IT flex.* | | | | -.019 | -.234 | 3.032 | |
| *Dynamism x Alignment DCs* | | | | .011 | .147 | 2.546 | |
| *Complexity x Alignment DCs* | | | | .010 | .099 | 4.429 | |
| *Hostility x Alignment DCs* | | | | -.024 | -.256 | 4.108 | |
| **Model D: Strategic alignment** | 322 | .300 | .282 | | | | 16.743 |
| *Size* | | | | .011 | .215 | 1.103 | |
| *Dynamism* | | | | -.059 | -1.124 | 1.246 | |
| *Complexity* | | | | -.049 | -.827 | 1.597 | |
| *Hostility* | | | | .208 | 3.627*** | 1.474 | |
| *Strategic Alignment* | | | | .495 | 9.423*** | 1.233 | |
| *Dynamism x Alignment* | | | | -.055 | -.942 | 1.524 | |
| *Complexity x Alignment* | | | | .034 | .505 | 2.004 | |
| *Hostility x Alignment* | | | | -.077 | -1.254 | 1.685 | |

Table 2.     *Regression analysis between independent variables (alignment), dependent variable (competitive performance), control and moderating variables and interaction terms.*
*\* p < 0.10, \*\* p < 0.001, \*\*\* p < 0.0001*

Also, alignment of IT flexibility dimensions explained a significant proportion of variance in competitive performance, i.e., $R^2 = .286$ (confirming hypothesis *1*). The same analysis was done for model B, and we found a similar result: $\beta = .463$, p< .0001, $R^2 = .278$ (confirming hypothesis *2*). Outcomes of simultaneous alignment (model C), i.e., the impact of two separate predictor variables, shows significant regression coefficient of our two main predictors (see Table 3) and has the highest explanatory power ($R^2 = .327$) of the variance in competitive performance and hence confirms hypothesis *3*.

Henceforth, analyses reveal various interesting findings. First, our main proposition that alignment of IT flexibility and dynamic capability dimensions has a positive impact on competitive firm performance, seems to be confirmed through the iterative testing of hypothesis 1-4a. Second, VIF scores are all within an acceptable range and thus reveal no multicollinearity problems. All F-scores indicate significant regressions suggesting the existence of a real effect of alignment as independent variables. Third, while 'strategic alignment' of underlying dimensions has the strongest impact on competitive performance (i.e., the highest regression coefficient of the main predictor, $\beta = .495$, p< .0001), 'simultaneous alignment' has the highest explanatory power ($R^2 = .327$). This basically contradicts our basic assumption (thereby rejecting hypothesis 4b) that 'strategic alignment' (single predictor variable covering all underlying IT flexibility and dynamic capability dimensions) would outperform any other alignment profiles in terms of impact and explanatory power; a particularity that must be discussed.

So, analyses suggest that alignment as a principal concept at the construct level first and foremost explains variance in competitive firm performance. This implies that our strategic alignment model can be a valuable tool to help firms to assess and improve their IT flexibility and dynamic capabilities. We controlled for the influence of various contextual (control) variables and moderation/interaction effects. Apparently, 'size' and the interaction effects do not seem to have any significant impact. As can be seen from Table 2 only 'Hostility' (main effect) seems to have a significant and positive relationship with competitive firm performance. This relation was added to model hostility as moderating effect and is ordinarily not interpreted as part of moderation. It could be the case that under conditions of high competition, firms need to increase their strategic alignment in order to enhance firm performance compared to competitors.

# 6 Discussion and conclusion

## 6.1 Findings and conclusions

Previous studies acknowledged the permeation of IS/IT in a wide variety of organizational aspects and highlighted the importance of synchronizing IT, organizational competences, and capabilities in enabling business abilities that help firms survive in the business ecosystem (Aral & Weill, 2007; Dao et al., 2011; Kim et al., 2011). IT flexibility and dynamic capabilities, in this respect, have been extensively researched in past work. Hence, we grounded this particular study on both the shortcomings and foundations of previous scholarship and focused on the question whether or not, strategic alignment of IT flexibility and dynamic capabilities contribute to higher levels of competitive performance. To this end, we conceptualized a strategic alignment model through a lens of the DCT and modular systems theory.

In doing so, we differentiate from prior research studies that test the direct or indirect but isolated impact of IT flexibility and (IT-enabled) dynamic capabilities on competitive performance (Byrd & Turner, 2000; Mikalef et al., 2016; Pavlou & El Sawy, 2011; Pil & Cohen, 2006; Tallon & Pinsonneault, 2011). In fact, we embrace an alignment approach (Henderson & Venkatraman, 1993; Scheper, 2002) that uses decompositions of the primary concepts (Overby et al., 2006). Outcomes suggest that simultaneous alignment of IT flexibility and dynamic capabilities (Model C) while controlling for contextual and interaction effects, has a significant impact (more than the alignment of IT flexibility or dynamic capabilities in isolation) on competitive firm performance (and the highest explanatory power). The fact that hypothesis 4b (strategic alignment outperforms other alignment models) was rejected might well be the result of a 'too broad' operationalized measure of strategic alignment whereby situational (case dependent) alignment, between two, separated, but interrelated con-structs are partially neglected. The 'strategic alignment' model (Model D) does outperform models A and B and has the highest regression coefficient. Contrary to other studies (Pavlou & El Sawy, 2006), we did not find significant interaction effects of the included moderation variables. The specific interpretation is currently still open on its own value. In sum, statistical analyses do show that achieving alignment comes with performance gains, including higher levels of financial performance, customer satisfaction, reduced operating costs, product and service quality and market share.

## 6.2    Toward alignment assessments in practice

This work has some interesting findings that can be applied in practice. The current paper reveals that strategic alignment is a potent source of value for IT/business managers and executives. We highlight the importance of strategically and situationally leveraging IT in core areas and the need for using both IT resources of high flexibility and developing capabilities for exploiting these resources, as a prerequisite for achieving superior performance. Consequently, this raises the need for managers to form multi-disciplinary networks of employees on all levels within the organization, including experts from both IT and the rest of the business functions, to deliver (on tactical and operational levels) the agreed objectives. This way, IT and business representatives are simultaneously engaged in the process of building, integrating, and reconfiguring internal and external IT and business competences and capabilities to address rapidly changing business ecosystem and allow a firm to adapt and reconfigure when needed. This relates well to the theses that flexible IT infrastructures will only be of value if leveraged appropriately to support or enable critical organizational capabilities that work towards dynamic strategic alignment (Chung, Rainer Jr, & Lewis, 2003). The strategic alignment model for IT flexibility and dynamic capabilities can guide decision-makers towards aligning the use of IT resources with their dynamic capabilities and for IS/IT and business investment to support the process of enhancing firm performance. The strategic alignment model can be applied in different situations and settings (i.e., generally applicable), hence enabling the measurement, monitoring, and comparison of IT flexibility and dynamic capabilities alignment through e.g., self-assessment in firms. Now, following our alignment models' logic, firms can define improvement activities that can be executed along the various dimension of our model under the following conditions: (1) investments meet a firms' current and future assessed needs; they are conditional on given situations such as the given state of alignment and the specified strategic alignment direction and (2) improvement activities are done simultaneously and hence by an integrated management perspective. In practice, firms define their own improvement roadmaps incrementally, radical or both as a strategy (Van de Wetering, Batenburg, & Lederman, 2010). In the course of the execution of all improvement activities, the level of alignment between the dimensions should be monitored so that synergies between improvement projects are developed. Critically reflecting on the chosen roadmap, or development path, while continuously maintaining alignment between the IT flexibility and dynamic capability dimensions and improve overall firm performance complements this process.

## 6.3    Limitations and venues for future research

This study has several limitations. These identified limitations are largely related to our data set because they were collected at a single point in time. First, a longitudinal approach would provide valuable insights into the evolutionary nature of alignment in time as for instance punctuated equilibrium models. We could investigate whether IS/IT alignment goes through long periods of relative stability, or evolutionary change, interrupted by short periods of quick and extensive, or revolutionary, change, as was demonstrated by Sabherwal et al. (Sabherwal et al., 2001). Second, our obtained survey data included various demographic variables (e.g., type, size, region), but our empirical analysis did not consider in-depth the possible differences among group segments and clusters. (Bergeron et al., 2004). Also, comparing alignment scores and outcomes across countries and groups might well contribute to the generalizability of our findings. A larger sample space would also then provide more robust results and enable cross-country comparisons as well as identification of differences between industries.

Third, despite significant findings of this study on a rather abstract level, we have not yet tested for potential differences between sample (sub)groups (and their interactions). Alignment might be conditioned to certain contextual and organizational elements. Potential future research can investigate and identify other exclusive configurational and contingency patterns and antecedents of alignment (Fiss, 2007; Mikalef, Pateli, Batenburg, & Van de Wetering, 2015) and cluster (Bergeron et al., 2004) that further contribute to high competitive performance. Outcomes could then possibly unfold mechanisms that limit or even condition strategic alignment. In order to address the above limitations, we are currently in the process of doing follow-up research. We hope that the current study gives rise to new research in this particular domain focusing on practical applications and tools that enable IT and business managers, and decision-makers to improve their firm's IT flexibility and dynamic capabilities, and ultimately increase their competitive firm performance.

## Acknowledgments

This project has received funding from the European Union's Horizon 2020 research and innovation programme under the Marie Sklodowska-Curie grant agreement No 704110.

## Appendix A: Survey items

|  | Mean | S.D. |
|---|---|---|
| **IT Flexibility** | | |
| *To what extent do you agree with the following statements? (1 – totally disagree 7 – totally agree)* | | |
| **- Modularity** | | |
| Our information systems are highly modular | 5.11 | 1.34 |
| The manner in which the components of our information systems are organized and integrated allows for rapid changes | 4.80 | 1.39 |
| Functionality can be quickly added to critical applications based on end-user requests | 4.78 | 1.50 |
| Exchanging or modifying single components does not affect our IT infrastructure | 4.89 | 1.56 |
| Organizational IT infrastructure and applications are developed on the basis of minimal unnecessary interdependencies | 4.65 | 1.45 |
| Organizational IT infrastructure and applications are loosely coupled | 4.59 | 1.50 |
| **- Transparency** | | |
| Remote users can seamlessly access centralized data and processes | 5.45 | 1.52 |
| Our user interfaces provide transparent access to all platforms and applications | 5.25 | 1.49 |
| Software applications can be easily transported and used across multiple platforms | 4.71 | 1.58 |
| Data of one system can be easily used in other systems | 4.77 | 1.56 |
| Our firm offers multiple interfaces or entry points (e.g., web access) to external users. | 4.89 | 1.49 |
| **- Standardization** | | |
| We have established corporate rules and standards for hardware and operating systems to ensure platform compatibility | 5.27 | 1.51 |
| We have identified and standardized data to be shared across systems and business units | 5.09 | 1.45 |
| Our systems are developed in order to incorporate electronic links to external parties | 4.78 | 1.69 |
| Organizational IT infrastructure and applications are highly interoperable | 5.29 | 1.50 |
| Organizational IT applications are developed based on compliance guidelines. | 5.16 | 1.58 |
| **- Scalability** | | |
| Our IT infrastructure easily compensates peaks in transaction volumes | 5.31 | 1.40 |
| Our information systems are scalable | 5.57 | 1.31 |
| Our IT infrastructure offers sufficient capacity in order to fulfill additional orders | 5.59 | 1.31 |
| The performance of our IT infrastructure completely fulfills our business needs regardless of usage magnitude | 5.25 | 1.46 |
| **IT-Enabled Dynamic Capabilities** | | |
| *Please indicate how effective your company is in using IT systems for the following purposes: (1-Not effective at all, 7-Highly effective)* | | |
| **- Sensing** | | |
| Scanning the environment and identifying new business opportunities | 4.88 | 1.51 |
| Reviewing our product development efforts to ensure they are in line with what the customers want | 5.06 | 1.36 |
| Implementing ideas for new products and improving existing products or services | 5.29 | 1.32 |
| Anticipating discontinuities arising in our business domain by developing greater reactive and proactive strength | 4.82 | 1.32 |
| **- Coordinating** | | |
| Providing more effective coordination among different functional activities | 5.12 | 1.35 |
| Providing more effective coordination with customers, business partners and distributors | 5.24 | 1.20 |
| Ensuring that the output of work is synchronized with the work of other functional units or business partners | 5.03 | 1.30 |
| Reducing redundant tasks, or overlapping activities performed by different operational units | 4.90 | 1.48 |
| **- Learning** | | |
| Identify, evaluate, and import new information and knowledge | 5.14 | 1.40 |
| Transform existing information into new knowledge | 5.01 | 1.34 |
| Assimilate new information and knowledge | 5.09 | 1.38 |
| Use accumulated information and knowledge to assist decision making | 5.08 | 1.35 |
| **- Integrating** | | |
| Easily accessing data and other valuable resources in real time from business partners | 4.92 | 1.43 |
| Aggregating relevant information from business partners, suppliers and customers. (e.g., operating information, business customer performance) | 4.99 | 1.37 |
| Collaborating in demand forecasting and planning between our firm and our business partners | 4.68 | 1.49 |
| Streamlining business processes with suppliers, distributors, and customers | 4.87 | 1.40 |
| **- Reconfiguring** | | |
| Adjusting for and responding to unexpected changes easily | 4.82 | 1.37 |
| Easily adding an eligible new partner that you want to do business with, or removing ones which you have terminated your partnership | 4.95 | 1.41 |
| Adjusting our business processes in response to shifts in our business priorities | 4.91 | 1.33 |
| Reconfiguring our business processes in order to come up with new productive assets | 4.74 | 1.40 |
| **Competitive Performance** | | |
| *Compared with your key competitors, please indicate how much you agree or disagree with the following statements regarding the degree to which you perform better than them: (1 – totally disagree 7 – totally agree)* | | |
| Return on investment (ROI) | 4.60 | 1.43 |
| Profits as percentage of sales | 4.57 | 1.40 |
| Decreasing product or service delivery cycle time | 4.57 | 1.51 |
| Rapid response to market demand | 4.77 | 1.65 |
| Rapid confirmation of customer orders | 4.87 | 1.58 |
| Increasing customer satisfaction | 5.07 | 1.58 |
| In profit growth rates | 4.54 | 1.48 |
| In reducing operating costs | 4.65 | 1.59 |
| Providing better product and service quality | 5.09 | 1.65 |
| Increasing our market share | 4.98 | 1.58 |